\documentclass[twocolumn,apl, showpacs,superscriptaddress]{revtex4}

\usepackage{bm}
\usepackage{amssymb}
\usepackage{amsmath}
\usepackage{graphics}
\usepackage{epsfig}
\usepackage{color}
\newcommand{\Cu}{\ensuremath{\mathrm{Cu}}}
\newcommand{\Ni}{\ensuremath{\mathrm{Ni}}}
\newcommand{\Nb}{\ensuremath{\mathrm{Nb}}}
\newcommand{\Al}{\ensuremath{\mathrm{Al}}}
\newcommand{\Si}{\ensuremath{\mathrm{Si}}}
\renewcommand{\O}{\ensuremath{\mathrm{O}}}

\newcommand{\q}[1]{#1}
\renewcommand{\section}[1]{}
\renewcommand{\subsection}[1]{}

\begin{document}


\title{%
  High quality ferromagnetic $0$ and $\pi$ Josephson tunnel junctions
}

\author{M. Weides}
\affiliation{%
  Center of Nanoelectronic Systems for Information Technology (CNI), Research Centre J\"ulich, D-52425 J\"ulich, Germany%
}

\author{M. Kemmler}
\affiliation{%
  Physikalisches Institut -- Experimentalphysik II,
  Universit\"at T\"ubingen,
  Auf der Morgenstelle 14,
  D-72076 T\"ubingen, Germany
}

\author{H. Kohlstedt}
\affiliation{%
  Center of Nanoelectronic Systems for Information Technology (CNI), Research Centre J\"ulich, D-52425 J\"ulich, Germany%
}
\affiliation{%
  Department of Material Science and Engineering and Department of Physics, University of Berkeley, California 94720, USA%
}

\author{A. Buzdin}
\affiliation{Institut Universitaire de France and Condensed Matter
Theory Group, CPMOH, University Bordeaux 1, UMR 5798, CNRS, F-33405
Talence Cedex, France}

\author{E. Goldobin}
\author{D. Koelle}
\author{R. Kleiner}
\affiliation{%
  Physikalisches Institut -- Experimentalphysik II,
  Universit\"at T\"ubingen,
  Auf der Morgenstelle 14,
  D-72076 T\"ubingen, Germany
}


\begin{abstract}
  We fabricated high quality $\Nb/\Al_2\O_3/\Ni_{0.6}\Cu_{0.4}/\Nb$ superconductor-insulator-ferromagnet-superconductor Josephson tunnel junctions. Depending on the thickness of the ferromagnetic $\Ni_{0.6}\Cu_{0.4}$ layer and on the ambient temperature, the junctions were in the $0$ or $\pi$ ground state. All junctions have homogeneous interfaces showing almost perfect Fraunhofer patterns. The $\Al_2\O_3$ tunnel barrier allows to achieve rather low damping, which is desired for many experiments especially in the quantum domain. The McCumber parameter $\beta_c$ increases exponentially with decreasing temperature and reaches $\beta_c\approx700$ at $T=2.11\,{\rm K}$. The critical current density in the $\pi$ state was up to $5\:\rm{A/cm^2}$ at $T=2.11\,{\rm K}$, resulting in a Josephson penetration depth $\lambda_J$ as low as $160\:\rm{\mu m}$. Experimentally determined junction parameters are well described by theory taking into account spin-flip scattering in the $\Ni_{0.6}\Cu_{0.4}$ layer and different transparencies of the interfaces.
\end{abstract}



\maketitle


The realization of solid state qubits attracts considerable interest. Josephson junctions (JJs) are used to realize
charge\cite{Nakamura}, phase\cite{Martinis:2002:RabiLJJ} or flux\cite{Mooij:1999:JosPersistCurrentQubit} qubits. For
the ``quiet'' flux qubit\cite{Ioffe:1999:sds-waveQubit}, which is self-biased and well decoupled from the environment,
one needs to use high quality $\pi$ JJs with high resistance (to avoid decoherence) and reasonably high critical
current density $j_c$ (to have the Josephson energy $E_J \gg k_BT$ for junction sizes of few microns or below). High
$j_c$ is also required to keep the Josephson plasma frequency $\omega_p \propto \sqrt{j_c}$, which plays the role of an
attempt frequency in the quantum tunneling problem, on the level of a few GHz.



The concept of $\pi$ JJs was introduced long ago\cite{Bulaevskii1977,Buzdin1991}, but only recently super\-con\-duc\-tor-fer\-ro\-mag\-net-su\-per\-con\-duc\-tor (SFS) $\pi$ JJs were realized\cite{RyazanovOboznovRusanoval.2001Couplingoftwosuperconductorsthroughaferromagnet:Evidenceforapijunction,Blum:2002:IcOscillations}.
Unfortunately SFS $\pi$ JJs are highly overdamped and cannot be used for applications where low dissipation is required. The obvious way to
decrease damping is to make a SFS-like \emph{tunnel} junction, i.e. a superconductor-insulator-fer\-ro\-mag\-net-su\-per\-con\-duc\-tor (SIFS) junction.
Due to the presence of the tunnel barrier the critical current $I_c$ in SIFS is lower than in SFS, but both the resistance $R$ (at $I \gtrsim I_c$)
and the $I_c R$ product are much higher. Moreover, the value of $I_c$ and $R$ can be tuned by changing the thickness $d_I$ of the insulator (tunnel
barrier).

A set of SIFS JJs with different thickness $d_F$ of the F-layer were recently fabricated and JJs with both $0$ and $\pi$ ground states were observed
depending on $d_F$\cite{Kontos:2002:SIFS-PiJJ}. Although, in the $\pi$ state the specific resistance of the barrier was high ($\rho\sim3\,{\rm
m\Omega\cdot cm^2}$), $j_c$ was below $7\:\rm{mA/cm^2}$ at $1.5\,{\rm K}$, resulting in an $I_c R$ product below $20\,{\rm \mu V}$, as can be
estimated from the data in Ref.~\onlinecite{Kontos:2002:SIFS-PiJJ}.

In this letter we report on fabrication and characterization of high quality Nb/AlO$_x$/Ni$_y$Cu$_{1-y}$/Nb JJs with different $d_F$ having as high as possible $j_c$ and $I_cR$ values. In the $\pi$ state we reached $j_c$ up to $5\,\mathrm{A/cm^2}$ at $T=2.11\,\mathrm{K}$ and maximum $I_c R$ values $\approx400\,\mathrm{\mu V}$. SIFS and reference SIS JJs were fabricated in-situ by magnetron sputtering and patterned using optical lithography and (reactive) dry-etching \cite{WeidesTillmannKohlstedtFabricationofhighqualityferromagneticJosephsonjunctions}. On thermally oxidized $\Si$ wafers we deposited $120\:\mathrm{nm}$ $\Nb$ and $5\:\mathrm{nm}$ $\Al$. To form the $\Al_2\O_3$ barrier (which should be as thin as possible, but without pinholes) we oxidized at $0.015$ or at $50\:\mathrm{mbar}$ to have $j_c^{(1)} \approx 4.0\:\mathrm{kA/cm^2}$ (wafer 1) and $j_c^{(2)}\approx0.19\:\mathrm{kA/cm^2}$ (wafer 2) for reference SIS JJs. For reference SIS JJs on wafer 1 the $I_c R$ product was $1.55\:\mathrm{mV}$.

To control the properties of SIFS JJs the thickness and the
roughness of the F-layer should be controlled on a sub-nm scale. To provide uniform growth of the F-layer, a $2\:\rm{nm}$ $\Cu$ interlayer was deposited between I-layer and F-layer. As F-layer we used diluted
$\Ni_{0.6}\Cu_{0.4}$, followed by a $40\:\mathrm{nm}$ $\Nb$
cap-layer. To produce JJs with different $d_F$ in a single run,
during sputtering of the F-layer, the substrate and sputter target were
shifted about half the substrate length producing a wedge-like
F-layer with $d_F$ from $1$ to $15\:\mathrm{nm}$ across the 4"
wafer. All other layers had uniform thicknesses. The SIFS junctions
had a squared shape with an area of $100\times 100\:\mathrm{\mu
m^2}$.


We have used diluted $\Ni_y\Cu_{1-y}$ alloy rather than pure $\Ni$ to have suitable $d_F$ (much larger than roughness) for the $\pi$ state. In \emph{very diluted} alloy with $y\leq 0.53$ strong spin-flip scattering \cite{SellierBaraducLeflochal.2003Temperature-inducedcrossoverbetween0andpistatesinS/F/Sjunctions} and $\Ni$ cluster formation are observed\cite{levin74,houghton70}. To avoid this magnetic inhomogeneity we have used $y=0.6$, as confirmed by Rutherford backscattering spectroscopy. The Curie temperature $T_\mathrm{C} \sim 225\:\mathrm{K}$ was determined by SQUID magnetometry and anisotropic Hall measurements on bare $\Ni_{0.6}\Cu_{0.4}$ films. Both $T_\mathrm{C}$ and resistivity $\rho_F(10\:\mathrm{K})=54\:\mathrm{\mu \Omega\cdot cm}$ are in good agreement with the literature\cite{hindmarchNi62Cu,ododoCriticalconcentrationforonsetofferromagnetisminCuNialloys}. The magnetization of such thin $\Ni_{0.6}\Cu_{0.4}$ films is in-plane. Interpolation of the magnetic moment $\mu$ from published data \cite{brouers73,RyazanovOboznovProkofieval.2004Superconductor-ferromagnet-superconductorpi-junctions,RusanovBoogaardHesselberthal.2002Inhomogeneoussuperconductivityinducedinaweakferromagnet,SellierBaraducLeflochal.2003Temperature-inducedcrossoverbetween0andpistatesinS/F/Sjunctions} yields $\mu=0.15\:\mu_B$ per atom for our $\Ni_{0.6}\Cu_{0.4}$ alloy.

\begin{figure}[b]
  \includegraphics{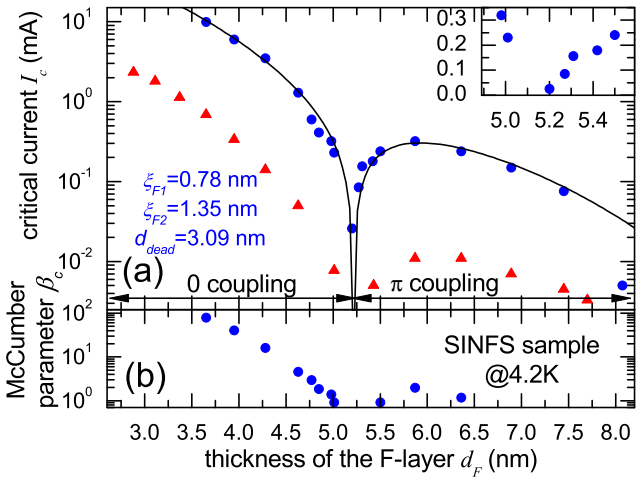}
  \caption{(Color online) $I_c(d_F)$ (a) and $\beta_c(d_F)$ (b) dependence (circles: wafer 1, triangles: wafer 2) and fitting curve for wafer 1. Inset shows magnification of $0$ to $\pi$ transition region for the wafer 1 on a linear scale.}
  \label{Fig:Ic(dF)}
\end{figure}

\begin{figure}[tb]
  \includegraphics{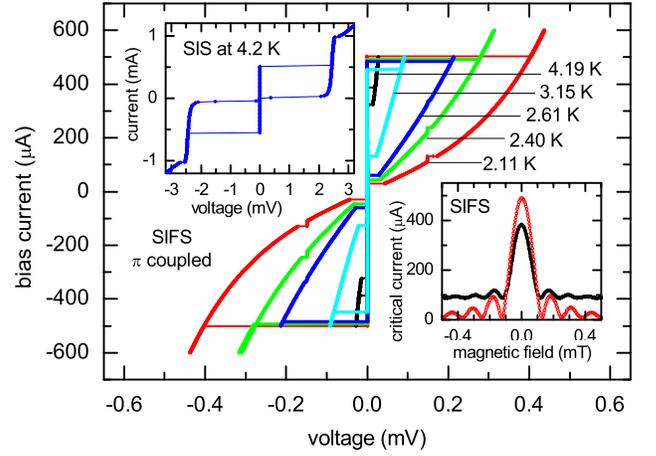}
  \caption{(Color online) IVCs of $\pi$ SIFS JJ ($d_F=5.87\:\mathrm{nm}$) at different $T$. Insets show: IVC of SIS JJ at $T=4.2\:\mathrm{K}$ (top) and $I_c(H)$ of SIFS JJ at $T=4.2$ and $2.11\:\mathrm{K}$}
  \label{IVIcBT}
\end{figure}

\begin{figure}[t]
  \includegraphics[width=8.6cm]{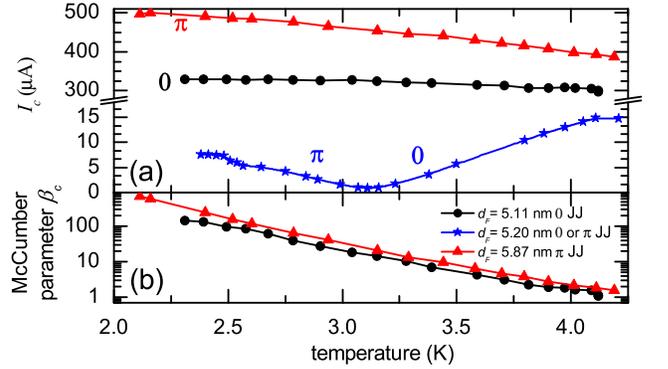}
  \caption{(Color online) $I_c(T)$ (a) and $\beta_c(T)$ (b) dependence of $0$ and $\pi$ SIFS JJs.}
  \label{Ic_T_beta}
\end{figure}

Following Ref.~\onlinecite{buzdinProximityeffectsinsuperconductor-ferromagnetheterostructures} one can derive that at $T \lesssim T_c$
\begin{equation}
  I_{c}(d_F) \sim \frac{1}{\gamma_{B2}}
  \exp\left( \frac{-d_F}{\xi_{F1}} \right)
  \cos \left( \frac{d_F-d_F^\mathrm{dead}}{\xi_{F2}} \right)
  ,\label{IcRn}
\end{equation}
where $\xi_{F1,F2}=\xi_F/\sqrt{\sqrt{1+\alpha^2}\pm\alpha}$ are the decay and oscillation lengths of order parameter\cite{DemlerArnoldBeasley1997Superconductingproximityeffectsinmagneticmetals}, $\xi_{F}=\sqrt{\hbar D/E_\mathrm{ex}}$ is the
decay/oscillation length without spin-flip scattering\cite{buzdinProximityeffectsinsuperconductor-ferromagnetheterostructures},
$E_\mathrm{ex}$ is the exchange energy, $\alpha=1/(\tau_s E_\mathrm{ex})$, $\tau_s$ is the
inelastic magnetic scattering
time\cite{RyazanovDouble-reversalthicknessdependenceofcriticalcurrentinsuperconductor-ferromagnet-superconductorJosephsonjunctions}
and $\gamma_{B2}$ is the transparency parameter of the SIF part treated like a single interface. \q{$d_F^\mathrm{dead}$ is the magnetic dead layer thickness.}
Eq.~(\ref{IcRn}) is derived assuming that the interfaces are not spin active,
cf.\cite{Cottet_Belzig_Superconducting_proximity_effect}, short decay length $\xi_{F1}<d_F$,
$\xi_{F1}\ll\xi_{F2}$ and FS interface transparency parameter $\gamma_{B1}=0$ ($\gamma_{B1}\ll\gamma_{B2}$). In comparison with \cite{BuzdinBaladieTheoreticaldescriptionofferromagneticpijunctionsnearthecriticaltemperature} Eq.~(\ref{IcRn}) takes into account magnetic impurity scattering which enters via $\tau_s$. Since $\xi_{F2}$ weakly depends on temperature $T$, the $0$-$\pi$ crossover can be observed by changing $T$.

The spread in $j_c$ among SIFS JJs with the same $d_F$ is about 2\% \cite{WeidesTillmannKohlstedtFabricationofhighqualityferromagneticJosephsonjunctions}. \q{The $I_c(d_F)$ dependence of our SIFS JJs is clearly non monotonic as shown in Fig.~\ref{Fig:Ic(dF)}. We argue that the minimum of $I_c(d_F)$ at $d_F\approx 5.21\,\mathrm{nm}$ corresponds to 0 to $\pi$ crossover. To rule out the possibility of 0-$\pi$ crossover at smaller $d_F$ we have investigated $I_c(d_F)$ down to $d_F=2\,\mathrm{nm}$ and did not observe any decrease or oscillation of $I_c(d_F)$. In Fig.~\ref{Fig:Ic(dF)} we show only data for ``low'' $j_c$ JJs ($L<2\lambda_J$) that we can treat as short JJs to fit experimental $I_c(d_F)$ using Eq.~(\ref{IcRn}). Due to a finite dead magnetic layer the change of phase takes place in an effectively reduced F-layer thickness. By fitting $I_c(d_F)$ for wafer 1 using Eq.~(\ref{IcRn}), we estimated $\xi_{F1}=0.78\:\mathrm{nm}$, $\xi_{F2}=1.35 \:\mathrm{nm}$ and $d_F^\mathrm{dead}=3.09\:\mathrm{nm}$. As we see, the inelastic magnetic scattering is strong ($\xi_{F1} < \xi_{F2}$) and the decay length $\xi_{F1} \ll d_F$, thus Eq.~(\ref{IcRn}) is applicable. Also, the found value of $d_F^\mathrm{dead}$ supports our claim that we observed 0 to $\pi$ rather than $\pi$ to 0 crossover.
}
According to Eq.(\ref{IcRn}) the coupling changes from $0$ to $\pi$ at the crossover thickness $d_{F}^{0\mbox{-}\pi} = \frac{\pi}{2}\xi_{F2}+d_F^\mathrm{dead}= 5.21\:\rm{nm}$, the shape of the $I_c(d_F)$ curve does not change with the thickness of the insulator, but the amplitude of $I_c(d_F)$ is proportional to the reciprocal transparency parameter $\gamma_{B2}^{-1}$. In the
interval of $d_F$ from $0$ (for SIS) to $9\,\mathrm{nm}$ the value of $j_c$ at $4.2\,\mathrm{K}$ changes over five orders of
magnitude from $4\:\mathrm{kA/cm^2}$ to below $0.05\:\mathrm{A/cm^2}$ (wafer 1).

The maximum $j_c$ in the $\pi$ state is $3.8\:\mathrm{A/cm^2}$ (wafer 1) and $j_c(\pi)=90\:\rm{mA/cm^2}$ (wafer 2) at $T=4.2\,\mathrm{K}$. This gives $\lambda_J \gtrsim 190\:\mathrm{\mu m}$, which can be easily \emph{increased} by increasing $d_I$. Further \emph{decrease} of $\lambda_J$ by lowering $d_I$ is limited by the appearance of microshorts in the barrier.

For comparison, in Ref.~\onlinecite{RyazanovDouble-reversalthicknessdependenceofcriticalcurrentinsuperconductor-ferromagnet-superconductorJosephsonjunctions} SFS JJs were fabricated using the weaker ferromagnet $\Ni_{0.53}\Cu_{0.47}$ ($T_\mathrm{C}=60\,\mathrm{K}$). Although the spin-flip scattering was also taken into account, the high interface transparencies ($\gamma_{B1}=0.52$) lead to a different $I_c(d_F)$ dependence than Eq.~(\ref{IcRn}) predicts. Also, the lower $E_\mathrm{ex}$ lead to larger $\xi_{F1}=1.24\:\mathrm{nm}$ and $\xi_{F2}=3.73\:\mathrm{nm}$. The magnetic dead layer was \q{$1.4$} times larger than in our system.

Fig.~\ref{Fig:Ic(dF)}(b) shows the dependence of the McCumber parameter $\beta_c(d_F)$, \q{which was estimated from the values of $I_c$ and $I_r$ (return current), at $T=4.2\:\mathrm{K}$ for wafer 1. The capacitance $C\approx800\,\mathrm{pF}$, determined from the Fiske step spacing of  $73\:\rm{\mu V}$ is nearly independent from $d_F$, but depends on $d_I$.} Near the $0$-$\pi$ crossover and for large $d_{F}$ the value of $I_c$ is very low and the junctions become
overdamped ($\beta_c < 0.7$). For $\pi$ JJs with $d_F$ near the maxima of the $I_c(d_{F})$ curve a hysteresis appears on the $I$-$V$ characteristic (IVC).

The IVCs and $I_c(H)$ patterns (voltage criterion $5\:\mathrm{\mu V}$) for a SIFS $\pi$ JJ with highest $I_c$ are shown in Fig.~\ref{IVIcBT}, c.f. the IVC of the SIS JJs shown in the inset. Theoretically, at lower temperature the quasiparticle current decreases and the gap appears at higher voltages. In experiment, due to heating effects at high bias currents, part of the sample became normal before we were able to reach the gap voltage. \q{At $T\leq2.61\:\rm{K}$ the first zero field step at $149\,\mathrm{\mu V}$ is visible on the IVC.}

The energy dependence of the density of states in $\Al$, $\Cu$ and $\Ni\Cu$ are not exactly BCS-like and $I_c(T)$ for SIFS JJs should show a more linear behavior \cite{RowellSmith1976Investigationofsuperconductingproximityeffectbyjosephsontunneling} than originally found by Ambegaokar-Baratoff \cite{AmbegaokarBaratoff1963Tunnelingbetweensuperconductors}. Variation of $T$ modifies $\xi_{F1}$ and $\xi_{F2}$ and can even change the ground state \cite{RyazanovOboznovRusanoval.2001Couplingoftwosuperconductorsthroughaferromagnet:Evidenceforapijunction,SellierBaraducLeflochal.2003Temperature-inducedcrossoverbetween0andpistatesinS/F/Sjunctions}. Since $E_\mathrm{ex}$ of $\Ni_{0.6}\Cu_{0.4}$ is relatively large, a change of $T$ affects our JJs much less than in previous work on the stronger diluted $\Ni\Cu$ alloys\cite{RyazanovOboznovRusanoval.2001Couplingoftwosuperconductorsthroughaferromagnet:Evidenceforapijunction,SellierBaraducLeflochal.2003Temperature-inducedcrossoverbetween0andpistatesinS/F/Sjunctions}. The $I_c(T)$ dependences for three JJs from wafer 1 are shown in Fig.~\ref{Ic_T_beta}(a). At $d_F=5.11\:\mathrm{nm}$ the JJ is $0$ coupled, but we attribute the nearly constant $I_c$ below $3.5\:\mathrm{K}$ to the interplay between increasing Cooper pair density and decreasing oscillation length $\xi_{F2}(T)$. The JJ with $d_F=5.20\:\mathrm{nm}$ is $0$ coupled at $T=4.2\:\mathrm{K}$, but changes coupling to $\pi$ below $3.11\:\rm{K}$. During the $0$--$\pi$ transition its critical current is not vanishing completely ($I_c^\mathrm{min}\approx0.8\,\mathrm{\mu A}$) either due to roughness of the ferromagnet or a prominent $\sin(2\phi)$ component in the current-phase relation\cite{Buzdin:2005:0-pi-trans,HouzetSinus2Phi}, which can appear intrinsically or again due to roughness\cite{Mints:2002:SplinteredVortices@GB,Buzdin:2003:phi-LJJ}. At the crossover temperature $T_x=3.11\,\mathrm{K}$, $I_c(H)$ can still be traced through several minima, so the large scale roughness must be small. The $d_F=5.87\:\mathrm{nm}$ JJ (also shown in Fig.~\ref{IVIcBT}) exhibits the highest critical current among $\pi$ JJs ($j_c=5\:\mathrm{A/cm^2}$ at $2.11\:\mathrm{K}$). Up to now the corresponding $\lambda_J=160\:\mathrm{\mu m}$ is the smallest achieved for SIFS JJs. Fig.~\ref{Ic_T_beta}(b) shows $\beta_c(T)$ for the same JJs. $\beta_c(T)$ increases exponentially below $4\,\mathrm{K}$ for both $0$ and $\pi$ JJs, indicating very weak Cooper pair breaking in the F-layer for these temperatures. The $\beta_c$ of the always overdamped JJ with $d_F=5.20\:\mathrm{nm}$ was not estimated.


In summary, we have fabricated and investigated SIFS Josephson junctions with $\Ni_{0.6}\Cu_{0.4}$ F-layer and thin $\Al_2\O_3$ tunnel barriers. The critical current $I_c$ changes sign as a function of the F-layer thickness $d_F$ in accordance with theory, exhibiting regions with $0$ and $\pi$ ground states. For $d_F$ near the $0$ to $\pi$ crossover the ground state can be controlled by changing the temperature. Our SIFS $\pi$ junctions show critical current densities $j_c$ up to $5\:\rm{A/cm^2}$ at $T=2.11\,{\rm K}$ and $I_c R$ products about $400\:\mathrm{\mu V}$. The achieved $\pi$ junction's Josephson penetration depth $\lambda_J$ as low as $160\:\rm{\mu m}$ at $2.11\,\mathrm{K}$ allows to fabricate long Josephson 0-$\pi$ junctions of reasonable size and study half integer flux quanta (semifluxons) that appear at the $0$-$\pi$ boundaries\cite{Kirtley:IcH-PiLJJ,Goldobin:SF-ReArrange,Susanto:SF-gamma_c} and have a size $\sim\lambda_J$. Reasonable $\lambda_J$ and low damping in such 0-$\pi$ junctions may lead to useful classical\cite{Lazarides:Ic(H):SF-Gen,Susanto:1D-Crystal} or quantum\cite{Kato:1997:QuTunnel0pi0JJ,Koyama:2005:d-dot:Qu,Goldobin:2005:QuTu2Semifluxons} circuits based on semifluxons.

We thank B. Hermanns for help with fabrication and V. Ryazanov for fruitful discussions. This work is supported by ESF program PiShift and by the DFG
projects GO-1106/1 and SFB/TR 21.


\begin{thebibliography}{37}
\expandafter\ifx\csname natexlab\endcsname\relax\def\natexlab#1{#1}\fi
\expandafter\ifx\csname bibnamefont\endcsname\relax
  \def\bibnamefont#1{#1}\fi
\expandafter\ifx\csname bibfnamefont\endcsname\relax
  \def\bibfnamefont#1{#1}\fi
\expandafter\ifx\csname citenamefont\endcsname\relax
  \def\citenamefont#1{#1}\fi
\expandafter\ifx\csname url\endcsname\relax
  \def\url#1{\texttt{#1}}\fi
\expandafter\ifx\csname urlprefix\endcsname\relax\def\urlprefix{URL }\fi
\providecommand{\bibinfo}[2]{#2}
\providecommand{\eprint}[2][]{\url{#2}}

\bibitem[{\citenamefont{Nakamura et~al.}(1999)\citenamefont{Nakamura, Pashkin,
  and Tsai}}]{Nakamura}
\bibinfo{author}{\bibfnamefont{Y.}~\bibnamefont{Nakamura}},
  \bibinfo{author}{\bibfnamefont{Y.~A.} \bibnamefont{Pashkin}},
  \bibnamefont{and} \bibinfo{author}{\bibfnamefont{J.~S.} \bibnamefont{Tsai}},
  \bibinfo{journal}{Nature} \textbf{\bibinfo{volume}{398}},
  \bibinfo{pages}{786} (\bibinfo{year}{1999}).

\bibitem[{\citenamefont{Martinis et~al.}(2002)\citenamefont{Martinis, Nam,
  Aumentado, and Urbina}}]{Martinis:2002:RabiLJJ}
\bibinfo{author}{\bibfnamefont{J.~M.} \bibnamefont{Martinis}},
  \bibinfo{author}{\bibfnamefont{S.}~\bibnamefont{Nam}},
  \bibinfo{author}{\bibfnamefont{J.}~\bibnamefont{Aumentado}},
  \bibnamefont{and} \bibinfo{author}{\bibfnamefont{C.}~\bibnamefont{Urbina}},
  \bibinfo{journal}{Phys. Rev. Lett.} \textbf{\bibinfo{volume}{89}},
  \bibinfo{eid}{117901} (\bibinfo{year}{2002}).

\bibitem[{\citenamefont{Mooij et~al.}(1999)\citenamefont{Mooij, Orlando,
  Levitov, Tian, van~der Wal, and Lloyd}}]{Mooij:1999:JosPersistCurrentQubit}
\bibinfo{author}{\bibfnamefont{J.~E.} \bibnamefont{Mooij}},
  \bibinfo{author}{\bibfnamefont{T.~P.} \bibnamefont{Orlando}},
  \bibinfo{author}{\bibfnamefont{L.}~\bibnamefont{Levitov}},
  \bibinfo{author}{\bibfnamefont{L.}~\bibnamefont{Tian}},
  \bibinfo{author}{\bibfnamefont{C.~H.} \bibnamefont{van~der Wal}},
  \bibnamefont{and} \bibinfo{author}{\bibfnamefont{S.}~\bibnamefont{Lloyd}},
  \bibinfo{journal}{Science} \textbf{\bibinfo{volume}{285}},
  \bibinfo{pages}{1036} (\bibinfo{year}{1999}).

\bibitem[{\citenamefont{Ioffe et~al.}(1999)\citenamefont{Ioffe, Geshkenbein,
  Feigel'man, Fauche\`ere, and Blatter}}]{Ioffe:1999:sds-waveQubit}
\bibinfo{author}{\bibfnamefont{L.~B.} \bibnamefont{Ioffe}},
  \bibinfo{author}{\bibfnamefont{V.~B.} \bibnamefont{Geshkenbein}},
  \bibinfo{author}{\bibfnamefont{M.~V.} \bibnamefont{Feigel'man}},
  \bibinfo{author}{\bibfnamefont{A.~L.} \bibnamefont{Fauche\`ere}},
  \bibnamefont{and} \bibinfo{author}{\bibfnamefont{G.}~\bibnamefont{Blatter}},
  \bibinfo{journal}{Nature} \textbf{\bibinfo{volume}{398}},
  \bibinfo{pages}{679} (\bibinfo{year}{1999}).

\bibitem[{\citenamefont{L.~Bulaevskii and Sobyanin}(1977)}]{Bulaevskii1977}
\bibinfo{author}{\bibfnamefont{V.~K.} \bibnamefont{L.~Bulaevskii}}
  \bibnamefont{and} \bibinfo{author}{\bibfnamefont{A.}~\bibnamefont{Sobyanin}},
  \bibinfo{journal}{JETP Lett.} \textbf{\bibinfo{volume}{25}},
  \bibinfo{pages}{7} (\bibinfo{year}{1977}).

\bibitem[{\citenamefont{Buzdin and Kupriyanov}(1991)}]{Buzdin1991}
\bibinfo{author}{\bibfnamefont{A.}~\bibnamefont{Buzdin}} \bibnamefont{and}
  \bibinfo{author}{\bibfnamefont{M.}~\bibnamefont{Kupriyanov}},
  \bibinfo{journal}{JETP Lett} \textbf{\bibinfo{volume}{53}},
  \bibinfo{pages}{321} (\bibinfo{year}{1991}).

\bibitem[{\citenamefont{Ryazanov et~al.}(2001)\citenamefont{Ryazanov, Oboznov,
  Rusanov, Veretennikov, Golubov, and
  Aarts}}]{RyazanovOboznovRusanoval.2001Couplingoftwosuperconductorsthroughafe%
rromagnet:Evidenceforapijunction}
\bibinfo{author}{\bibfnamefont{V.~V.} \bibnamefont{Ryazanov}},
  \bibinfo{author}{\bibfnamefont{V.~A.} \bibnamefont{Oboznov}},
  \bibinfo{author}{\bibfnamefont{A.~Y.} \bibnamefont{Rusanov}},
  \bibinfo{author}{\bibfnamefont{A.~V.} \bibnamefont{Veretennikov}},
  \bibinfo{author}{\bibfnamefont{A.~A.} \bibnamefont{Golubov}},
  \bibnamefont{and} \bibinfo{author}{\bibfnamefont{J.}~\bibnamefont{Aarts}},
  \bibinfo{journal}{Phys. Rev. Lett.} \textbf{\bibinfo{volume}{86}},
  \bibinfo{pages}{2427} (\bibinfo{year}{2001}).

\bibitem[{\citenamefont{Blum et~al.}(2002)\citenamefont{Blum, Tsukernik,
  Karpovski, and Palevski}}]{Blum:2002:IcOscillations}
\bibinfo{author}{\bibfnamefont{Y.}~\bibnamefont{Blum}},
  \bibinfo{author}{\bibfnamefont{A.}~\bibnamefont{Tsukernik}},
  \bibinfo{author}{\bibfnamefont{M.}~\bibnamefont{Karpovski}},
  \bibnamefont{and} \bibinfo{author}{\bibfnamefont{A.}~\bibnamefont{Palevski}},
  \bibinfo{journal}{Phys. Rev. Lett.} \textbf{\bibinfo{volume}{89}},
  \bibinfo{pages}{187004} (\bibinfo{year}{2002}).

\bibitem[{\citenamefont{Kontos et~al.}(2002)\citenamefont{Kontos, Aprili,
  Lesueur, Gen\^et, Stephanidis, and Boursier}}]{Kontos:2002:SIFS-PiJJ}
\bibinfo{author}{\bibfnamefont{T.}~\bibnamefont{Kontos}},
  \bibinfo{author}{\bibfnamefont{M.}~\bibnamefont{Aprili}},
  \bibinfo{author}{\bibfnamefont{J.}~\bibnamefont{Lesueur}},
  \bibinfo{author}{\bibfnamefont{F.}~\bibnamefont{Gen\^et}},
  \bibinfo{author}{\bibfnamefont{B.}~\bibnamefont{Stephanidis}},
  \bibnamefont{and} \bibinfo{author}{\bibfnamefont{R.}~\bibnamefont{Boursier}},
  \bibinfo{journal}{Phys. Rev. Lett.} \textbf{\bibinfo{volume}{89}},
  \bibinfo{pages}{137007} (\bibinfo{year}{2002}).

\bibitem[{\citenamefont{Weides et~al.}(2006)\citenamefont{Weides, Tillmann, and
  Kohlstedt}}]{WeidesTillmannKohlstedtFabricationofhighqualityferromagneticJos%
ephsonjunctions}
\bibinfo{author}{\bibfnamefont{M.}~\bibnamefont{Weides}},
  \bibinfo{author}{\bibfnamefont{K.}~\bibnamefont{Tillmann}}, \bibnamefont{and}
  \bibinfo{author}{\bibfnamefont{H.}~\bibnamefont{Kohlstedt}},
  \bibinfo{journal}{Physica C} \textbf{\bibinfo{volume}{437-438}},
  \bibinfo{pages}{349} (\bibinfo{year}{2006}).

\bibitem[{\citenamefont{Sellier et~al.}(2003)\citenamefont{Sellier, Baraduc,
  Lefloch, and
  Calemczuk}}]{SellierBaraducLeflochal.2003Temperature-inducedcrossoverbetween%
0andpistatesinS/F/Sjunctions}
\bibinfo{author}{\bibfnamefont{H.}~\bibnamefont{Sellier}},
  \bibinfo{author}{\bibfnamefont{C.}~\bibnamefont{Baraduc}},
  \bibinfo{author}{\bibfnamefont{F.}~\bibnamefont{Lefloch}}, \bibnamefont{and}
  \bibinfo{author}{\bibfnamefont{R.}~\bibnamefont{Calemczuk}},
  \bibinfo{journal}{Phys. Rev. B} \textbf{\bibinfo{volume}{68}},
  \bibinfo{pages}{054531} (\bibinfo{year}{2003}).

\bibitem[{\citenamefont{Levin and Mills}(1974)}]{levin74}
\bibinfo{author}{\bibfnamefont{K.}~\bibnamefont{Levin}} \bibnamefont{and}
  \bibinfo{author}{\bibfnamefont{D.~L.} \bibnamefont{Mills}},
  \bibinfo{journal}{Phys. Rev. B} \textbf{\bibinfo{volume}{9}},
  \bibinfo{pages}{2354} (\bibinfo{year}{1974}).

\bibitem[{\citenamefont{Houghton et~al.}(1970)\citenamefont{Houghton, Sarachik,
  and Kouvel}}]{houghton70}
\bibinfo{author}{\bibfnamefont{R.~W.} \bibnamefont{Houghton}},
  \bibinfo{author}{\bibfnamefont{M.~P.} \bibnamefont{Sarachik}},
  \bibnamefont{and} \bibinfo{author}{\bibfnamefont{J.~S.}
  \bibnamefont{Kouvel}}, \bibinfo{journal}{Phys. Rev. Lett.}
  \textbf{\bibinfo{volume}{25}}, \bibinfo{pages}{238} (\bibinfo{year}{1970}).

\bibitem[{\citenamefont{Hindmarch et~al.}(2005)\citenamefont{Hindmarch,
  Marrows, and Hickey}}]{hindmarchNi62Cu}
\bibinfo{author}{\bibfnamefont{A.~T.} \bibnamefont{Hindmarch}},
  \bibinfo{author}{\bibfnamefont{C.~H.} \bibnamefont{Marrows}},
  \bibnamefont{and} \bibinfo{author}{\bibfnamefont{B.~J.}
  \bibnamefont{Hickey}}, \bibinfo{journal}{Phys. Rev. B}
  \textbf{\bibinfo{volume}{72}}, \bibinfo{pages}{100401}
  (\bibinfo{year}{2005}).

\bibitem[{\citenamefont{Ododo and
  Coles}(1977)}]{ododoCriticalconcentrationforonsetofferromagnetisminCuNialloy%
s}
\bibinfo{author}{\bibfnamefont{J.~C.} \bibnamefont{Ododo}} \bibnamefont{and}
  \bibinfo{author}{\bibfnamefont{B.~R.} \bibnamefont{Coles}},
  \bibinfo{journal}{J. Phys. F: Met. Phys.} \textbf{\bibinfo{volume}{7}},
  \bibinfo{pages}{2393} (\bibinfo{year}{1977}).

\bibitem[{\citenamefont{Brouers et~al.}(1973)\citenamefont{Brouers, Vedyayev,
  and Giorgino}}]{brouers73}
\bibinfo{author}{\bibfnamefont{F.}~\bibnamefont{Brouers}},
  \bibinfo{author}{\bibfnamefont{A.~V.} \bibnamefont{Vedyayev}},
  \bibnamefont{and} \bibinfo{author}{\bibfnamefont{M.}~\bibnamefont{Giorgino}},
  \bibinfo{journal}{Phys. Rev. B} \textbf{\bibinfo{volume}{7}},
  \bibinfo{pages}{380} (\bibinfo{year}{1973}).

\bibitem[{\citenamefont{Ryazanov et~al.}(2004)\citenamefont{Ryazanov, Oboznov,
  Prokofiev, Bolginov, and
  A.K.Feofanov}}]{RyazanovOboznovProkofieval.2004Superconductor-ferromagnet-su%
perconductorpi-junctions}
\bibinfo{author}{\bibfnamefont{V.~V.} \bibnamefont{Ryazanov}},
  \bibinfo{author}{\bibfnamefont{V.~A.} \bibnamefont{Oboznov}},
  \bibinfo{author}{\bibfnamefont{A.~S.} \bibnamefont{Prokofiev}},
  \bibinfo{author}{\bibfnamefont{V.}~\bibnamefont{Bolginov}}, \bibnamefont{and}
  \bibinfo{author}{\bibnamefont{A.K.Feofanov}}, \bibinfo{journal}{J. Low. Temp.
  Phys.} \textbf{\bibinfo{volume}{136}}, \bibinfo{pages}{385}
  (\bibinfo{year}{2004}).

\bibitem[{\citenamefont{Rusanov et~al.}(2002)\citenamefont{Rusanov, Boogaard,
  Hesselberth, Seiler, and
  Aarts}}]{RusanovBoogaardHesselberthal.2002Inhomogeneoussuperconductivityindu%
cedinaweakferromagnet}
\bibinfo{author}{\bibfnamefont{A.}~\bibnamefont{Rusanov}},
  \bibinfo{author}{\bibfnamefont{R.}~\bibnamefont{Boogaard}},
  \bibinfo{author}{\bibfnamefont{M.}~\bibnamefont{Hesselberth}},
  \bibinfo{author}{\bibfnamefont{H.}~\bibnamefont{Seiler}}, \bibnamefont{and}
  \bibinfo{author}{\bibfnamefont{J.}~\bibnamefont{Aarts}},
  \bibinfo{journal}{Physica C} \textbf{\bibinfo{volume}{369}},
  \bibinfo{pages}{300} (\bibinfo{year}{2002}).

\bibitem[{\citenamefont{Buzdin}(2005{\natexlab{a}})}]{buzdinProximityeffectsin%
superconductor-ferromagnetheterostructures}
\bibinfo{author}{\bibfnamefont{A.~I.} \bibnamefont{Buzdin}},
  \bibinfo{journal}{Rev. Mod. Phys.} \textbf{\bibinfo{volume}{77}},
  \bibinfo{pages}{935} (\bibinfo{year}{2005}{\natexlab{a}}).

\bibitem[{\citenamefont{Demler et~al.}(1997)\citenamefont{Demler, Arnold, and
  Beasley}}]{DemlerArnoldBeasley1997Superconductingproximityeffectsinmagneticm%
etals}
\bibinfo{author}{\bibfnamefont{E.~A.} \bibnamefont{Demler}},
  \bibinfo{author}{\bibfnamefont{G.~B.} \bibnamefont{Arnold}},
  \bibnamefont{and} \bibinfo{author}{\bibfnamefont{M.~R.}
  \bibnamefont{Beasley}}, \bibinfo{journal}{Phys. Rev. B}
  \textbf{\bibinfo{volume}{55}}, \bibinfo{pages}{15174} (\bibinfo{year}{1997}).

\bibitem[{\citenamefont{Oboznov et~al.}(2006)\citenamefont{Oboznov, Bol'ginov,
  Feofanov, Ryazanov, and
  Buzdin}}]{RyazanovDouble-reversalthicknessdependenceofcriticalcurrentinsuper%
conductor-ferromagnet-superconductorJosephsonjunctions}
\bibinfo{author}{\bibfnamefont{V.~A.} \bibnamefont{Oboznov}},
  \bibinfo{author}{\bibfnamefont{V.~V.} \bibnamefont{Bol'ginov}},
  \bibinfo{author}{\bibfnamefont{A.~K.} \bibnamefont{Feofanov}},
  \bibinfo{author}{\bibfnamefont{V.~V.} \bibnamefont{Ryazanov}},
  \bibnamefont{and} \bibinfo{author}{\bibfnamefont{A.~I.}
  \bibnamefont{Buzdin}}, \bibinfo{journal}{Phys. Rev. Lett.}
  \textbf{\bibinfo{volume}{96}}, \bibinfo{pages}{197003}
  (\bibinfo{year}{2006}).

\bibitem[{\citenamefont{Cottet and
  Belzig}(2005)}]{Cottet_Belzig_Superconducting_proximity_effect}
\bibinfo{author}{\bibfnamefont{A.}~\bibnamefont{Cottet}} \bibnamefont{and}
  \bibinfo{author}{\bibfnamefont{W.}~\bibnamefont{Belzig}},
  \bibinfo{journal}{Phys. Rev. B} \textbf{\bibinfo{volume}{72}},
  \bibinfo{pages}{180503} (\bibinfo{year}{2005}).

\bibitem[{\citenamefont{Buzdin and
  Baladie}(2003)}]{BuzdinBaladieTheoreticaldescriptionofferromagneticpijunctio%
nsnearthecriticaltemperature}
\bibinfo{author}{\bibfnamefont{A.}~\bibnamefont{Buzdin}} \bibnamefont{and}
  \bibinfo{author}{\bibfnamefont{I.}~\bibnamefont{Baladie}},
  \bibinfo{journal}{Phys. Rev. B} \textbf{\bibinfo{volume}{67}},
  \bibinfo{pages}{184519} (\bibinfo{year}{2003}).

\bibitem[{\citenamefont{Rowell and
  Smith}(1976)}]{RowellSmith1976Investigationofsuperconductingproximityeffectb%
yjosephsontunneling}
\bibinfo{author}{\bibfnamefont{N.~L.} \bibnamefont{Rowell}} \bibnamefont{and}
  \bibinfo{author}{\bibfnamefont{H.~J.~T.} \bibnamefont{Smith}},
  \bibinfo{journal}{Canadian J. Phys.} \textbf{\bibinfo{volume}{54}},
  \bibinfo{pages}{223} (\bibinfo{year}{1976}).

\bibitem[{\citenamefont{Ambegaokar and
  Baratoff}(1963)}]{AmbegaokarBaratoff1963Tunnelingbetweensuperconductors}
\bibinfo{author}{\bibfnamefont{V.}~\bibnamefont{Ambegaokar}} \bibnamefont{and}
  \bibinfo{author}{\bibfnamefont{A.}~\bibnamefont{Baratoff}},
  \bibinfo{journal}{Phys. Rev. Lett.} \textbf{\bibinfo{volume}{10}},
  \bibinfo{pages}{486} (\bibinfo{year}{1963}).

\bibitem[{\citenamefont{Buzdin}(2005{\natexlab{b}})}]{Buzdin:2005:0-pi-trans}
\bibinfo{author}{\bibfnamefont{A.}~\bibnamefont{Buzdin}},
  \bibinfo{journal}{Phys. Rev. B} \textbf{\bibinfo{volume}{72}},
  \bibinfo{eid}{100501} (\bibinfo{year}{2005}{\natexlab{b}}).

\bibitem[{\citenamefont{Houzet et~al.}(2005)\citenamefont{Houzet, Vinokur, and
  Pistolesi}}]{HouzetSinus2Phi}
\bibinfo{author}{\bibfnamefont{M.}~\bibnamefont{Houzet}},
  \bibinfo{author}{\bibfnamefont{V.}~\bibnamefont{Vinokur}}, \bibnamefont{and}
  \bibinfo{author}{\bibfnamefont{F.}~\bibnamefont{Pistolesi}},
  \bibinfo{journal}{Phys. Rev. B} \textbf{\bibinfo{volume}{72}},
  \bibinfo{pages}{220506} (\bibinfo{year}{2005}).

\bibitem[{\citenamefont{Mints et~al.}(2002)\citenamefont{Mints, Papiashvili,
  Kirtley, Hilgenkamp, Hammerl, and
  Mannhart}}]{Mints:2002:SplinteredVortices@GB}
\bibinfo{author}{\bibfnamefont{R.~G.} \bibnamefont{Mints}},
  \bibinfo{author}{\bibfnamefont{I.}~\bibnamefont{Papiashvili}},
  \bibinfo{author}{\bibfnamefont{J.~R.} \bibnamefont{Kirtley}},
  \bibinfo{author}{\bibfnamefont{H.}~\bibnamefont{Hilgenkamp}},
  \bibinfo{author}{\bibfnamefont{G.}~\bibnamefont{Hammerl}}, \bibnamefont{and}
  \bibinfo{author}{\bibfnamefont{J.}~\bibnamefont{Mannhart}},
  \bibinfo{journal}{Phys. Rev. Lett.} \textbf{\bibinfo{volume}{89}},
  \bibinfo{pages}{067004} (\bibinfo{year}{2002}).

\bibitem[{\citenamefont{Buzdin and Koshelev}(2003)}]{Buzdin:2003:phi-LJJ}
\bibinfo{author}{\bibfnamefont{A.}~\bibnamefont{Buzdin}} \bibnamefont{and}
  \bibinfo{author}{\bibfnamefont{A.~E.} \bibnamefont{Koshelev}},
  \bibinfo{journal}{Phys. Rev. B} \textbf{\bibinfo{volume}{67}},
  \bibinfo{eid}{220504} (\bibinfo{year}{2003}).

\bibitem[{\citenamefont{Kirtley et~al.}(1997)\citenamefont{Kirtley, Moler, and
  Scalapino}}]{Kirtley:IcH-PiLJJ}
\bibinfo{author}{\bibfnamefont{J.~R.} \bibnamefont{Kirtley}},
  \bibinfo{author}{\bibfnamefont{K.~A.} \bibnamefont{Moler}}, \bibnamefont{and}
  \bibinfo{author}{\bibfnamefont{D.~J.} \bibnamefont{Scalapino}},
  \bibinfo{journal}{Phys. Rev. B} \textbf{\bibinfo{volume}{56}},
  \bibinfo{pages}{886} (\bibinfo{year}{1997}).

\bibitem[{\citenamefont{Goldobin et~al.}(2003)\citenamefont{Goldobin, Koelle,
  and Kleiner}}]{Goldobin:SF-ReArrange}
\bibinfo{author}{\bibfnamefont{E.}~\bibnamefont{Goldobin}},
  \bibinfo{author}{\bibfnamefont{D.}~\bibnamefont{Koelle}}, \bibnamefont{and}
  \bibinfo{author}{\bibfnamefont{R.}~\bibnamefont{Kleiner}},
  \bibinfo{journal}{Phys. Rev. B} \textbf{\bibinfo{volume}{67}},
  \bibinfo{pages}{224515} (\bibinfo{year}{2003}).

\bibitem[{\citenamefont{Susanto et~al.}(2003)\citenamefont{Susanto, van Gils,
  Visser, Ariando, Smilde, and Hilgenkamp}}]{Susanto:SF-gamma_c}
\bibinfo{author}{\bibfnamefont{H.}~\bibnamefont{Susanto}},
  \bibinfo{author}{\bibfnamefont{S.~A.} \bibnamefont{van Gils}},
  \bibinfo{author}{\bibfnamefont{T.~P.~P.} \bibnamefont{Visser}},
  \bibinfo{author}{\bibnamefont{Ariando}},
  \bibinfo{author}{\bibfnamefont{H.-J.~H.} \bibnamefont{Smilde}},
  \bibnamefont{and}
  \bibinfo{author}{\bibfnamefont{H.}~\bibnamefont{Hilgenkamp}},
  \bibinfo{journal}{Phys. Rev. B} \textbf{\bibinfo{volume}{68}},
  \bibinfo{pages}{104501} (\bibinfo{year}{2003}).

\bibitem[{\citenamefont{Lazarides}(2004)}]{Lazarides:Ic(H):SF-Gen}
\bibinfo{author}{\bibfnamefont{N.}~\bibnamefont{Lazarides}},
  \bibinfo{journal}{Phys. Rev. B} \textbf{\bibinfo{volume}{69}},
  \bibinfo{eid}{212501} (\bibinfo{year}{2004}).

\bibitem[{\citenamefont{Susanto et~al.}(2005)\citenamefont{Susanto, Goldobin,
  Koelle, Kleiner, and van Gils}}]{Susanto:1D-Crystal}
\bibinfo{author}{\bibfnamefont{H.}~\bibnamefont{Susanto}},
  \bibinfo{author}{\bibfnamefont{E.}~\bibnamefont{Goldobin}},
  \bibinfo{author}{\bibfnamefont{D.}~\bibnamefont{Koelle}},
  \bibinfo{author}{\bibfnamefont{R.}~\bibnamefont{Kleiner}}, \bibnamefont{and}
  \bibinfo{author}{\bibfnamefont{S.~A.} \bibnamefont{van Gils}},
  \bibinfo{journal}{Phys. Rev. B} \textbf{\bibinfo{volume}{71}},
  \bibinfo{eid}{174510} (\bibinfo{year}{2005}).

\bibitem[{\citenamefont{Kato and Imada}(1997)}]{Kato:1997:QuTunnel0pi0JJ}
\bibinfo{author}{\bibfnamefont{T.}~\bibnamefont{Kato}} \bibnamefont{and}
  \bibinfo{author}{\bibfnamefont{M.}~\bibnamefont{Imada}}, \bibinfo{journal}{J.
  Phys. Soc. Jpn.} \textbf{\bibinfo{volume}{66}}, \bibinfo{pages}{1445}
  (\bibinfo{year}{1997}).

\bibitem[{\citenamefont{Koyama et~al.}(2005)\citenamefont{Koyama, Machida,
  Kato, and Ishida}}]{Koyama:2005:d-dot:Qu}
\bibinfo{author}{\bibfnamefont{T.}~\bibnamefont{Koyama}},
  \bibinfo{author}{\bibfnamefont{M.}~\bibnamefont{Machida}},
  \bibinfo{author}{\bibfnamefont{M.}~\bibnamefont{Kato}}, \bibnamefont{and}
  \bibinfo{author}{\bibfnamefont{T.}~\bibnamefont{Ishida}},
  \bibinfo{journal}{Physica C} \textbf{\bibinfo{volume}{426--431}},
  \bibinfo{pages}{1561} (\bibinfo{year}{2005}).

\bibitem[{\citenamefont{Goldobin et~al.}(2005)\citenamefont{Goldobin, Vogel,
  Crasser, Walser, Schleich, Koelle, and
  Kleiner}}]{Goldobin:2005:QuTu2Semifluxons}
\bibinfo{author}{\bibfnamefont{E.}~\bibnamefont{Goldobin}},
  \bibinfo{author}{\bibfnamefont{K.}~\bibnamefont{Vogel}},
  \bibinfo{author}{\bibfnamefont{O.}~\bibnamefont{Crasser}},
  \bibinfo{author}{\bibfnamefont{R.}~\bibnamefont{Walser}},
  \bibinfo{author}{\bibfnamefont{W.~P.} \bibnamefont{Schleich}},
  \bibinfo{author}{\bibfnamefont{D.}~\bibnamefont{Koelle}}, \bibnamefont{and}
  \bibinfo{author}{\bibfnamefont{R.}~\bibnamefont{Kleiner}},
  \bibinfo{journal}{Phys. Rev. B} \textbf{\bibinfo{volume}{72}},
  \bibinfo{eid}{054527} (\bibinfo{year}{2005}).

\end{thebibliography}

\end{document}